# Magnetic angular position sensor enabled by spin-orbit torque

Ziyan Luo, Yanjun Xu, Yumeng Yang, and Yihong Wu[a]

*Department of Electrical and Computer Engineering, National University of Singapore, 4 Engineering Drive 3, Singapore 117583, Singapore*

We propose a simple scheme for magnetic angular position sensor based on current-induced spin-orbit torque effect. A full range detection of 360º is realized with a pair of Hall crosses made of heavy metal/ferromagnet heterostructures. The current axes of the two Hall crosses are aligned orthogonal to each other such that when both devices are subject to a rotational in-plane magnetic field, the differential Hall voltage due to current pulses of opposite polarity exhibits a sine and cosine angular dependence on the field direction, respectively. The field rotational angle is then calculated from the sine and cosine output signals via the arctan2 function. A linear correspondence between the calculated and actual field angle is obtained in the field range of 500 – 2000 Oe, with an average angle error of 0.38 – 0.65º.

---

[a] Author to whom correspondence should be addressed: elewuyh@nus.edu.sg

Angular position sensor is an essential component for rotational motion control in a wide range of applications including automotive, consumer products, manufacturing, *etc*.[1-2] Among the different types of angular position sensors, magnetic sensors are mostly widely used because of their combined features of contactless detection, robust performance and low cost. A typical magnetic angle detection setup consists of a magnet fixed on a rotary shaft to generate a time varying magnetic flux and a stationary magnetic sensor to detect the flux[3]. Depending on whether it detects the time derivative, magnitude or direction of the magnetic flux, the sensor can be categorized into three major types, *i.e.*, inductive, Hall effect and magnetoresistance (MR) sensors. Each type of sensor has its own advantages and drawbacks. For example, the inductive sensor is robust and highly resistant to harsh environment, but at the same time, it tends to be bulky and more expensive as compared to Hall and MR sensors. Hall sensors, which are currently dominant in automotive applications[4-5], are low cost, but they have a moderate sensitivity at low field and a relatively large thermal drift. The MR sensors, including anisotropic magnetoresistance (AMR) sensor[6-8], giant magnetoresistance (GMR) sensor and tunneling magnetoresistance (TMR) sensor[9-10], fill up nicely the gap between Hall and inductive sensors in terms of field sensitivity, detection accuracy and cost. The MR sensors have been developed as read sensors for data storage applications. However, when properly configured, these sensors can produce a sinusoidal output signal with respect to the external field direction at saturate state; and therefore, they function naturally as an angular position sensor. Although the operation principle is simple, dedicated designs and fabrication processes are required in order to obtain distortion-free sinusoidal output signals; and therefore, they tend to be more expensive than Hall sensors. Here we report a very simple angular position sensor based on the recently discovered spin-orbit torque (SOT) effect[11]. The sensor exhibits the same sinusoidal angular dependence as the TMR and GMR sensors do, but it is structurally much simpler, and thus can potentially lead to an angular position sensor with a much lower cost as compared to its GMR and TMR counterparts.

The SOT in heavy metal (HM) / ferromagnet (FM) heterostructure has been investigated extensively in recent years as a promising mechanism for magnetization switching and related applications. Although the exact mechanism is still being debated, it is generally accepted that two types of torques are present in the HM/FM heterostructures, one is called field-like (FL) and the other is (anti)damping-like (DL). Phenomenally, the two types of torques can be modelled by $\vec{T}_{DL} = \tau_{DL}\hat{m} \times [\hat{m} \times (\hat{z} \times \hat{j})]$ and $\vec{T}_{FL} = \tau_{FL}\hat{m} \times (\hat{z} \times \hat{j})$, respectively, where $\hat{m}$ is the magnetization direction, $\hat{j}$ is the in-plane current density, $\hat{z}$ is the interface normal, and $\tau_{FL}$ and $\tau_{DL}$ are the magnitude of FL and DL torques, respectively[12-14]. If $\hat{m}$ does not change significantly, the two torques can be expressed in the form of $\vec{M} \times \vec{H}_{eff}$, where $\vec{H}_{eff}$ is an effective field. Following this notion, the FL effective field ($\vec{H}_{FL}$) is in the direction of $\hat{z} \times \hat{j}$, whereas the DL effective field ($\vec{H}_{DL}$) is in the direction of $\hat{m} \times (\hat{z} \times \hat{j})$ (note: the sign can be different depending on the stacking order of HM and FM with respect to the coordinate axes). In this work, we employed Pt as the HM and Co as the FM layer. In order to configure it for angular position sensing applications, we choose a Co thickness such that it will exhibit an in-plane magnetic anisotropy. Fig. 1(a) illustrates the directions of $\vec{H}_{DL}$ and $\vec{H}_{FL}$ in Pt/Co when current is in *x*-direction and magnetization is in the *xy* plane. Also shown is the Oersted field ($\vec{H}_{Oe}$) which in this case is in the same direction of $\vec{H}_{FL}$. As can be seen, $\vec{H}_{DL}$ points in *z*-direction, which causes an out-of-plane tilt of the magnetization, and thereby gives rise to an anomalous Hall effect (AHE) signal. On the other hand, $\vec{H}_{FL}$ and $\vec{H}_{Oe}$ rotate the magnetization in the sample plane, generating a planar Hall effect (PHE) signal[15-17].

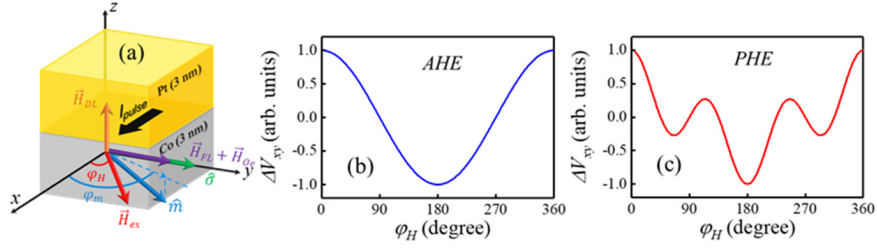

FIG. 1. (a) Illustration of the SOT effective fields in a Pt/Co bilayer induced by a current in *x*-direction ($\vec{H}_{ex}$: external field, $\hat{m}$: magnetization direction, $\hat{\sigma}$: spin polarization). (b) and (c): simulated differential AHE (b) and PHE (c) signals as a function of $\varphi_H$ induced by two current pulses with opposite polarity.

The change of AHE and PHE signal induced by the SOT effective fields can be derived analytically by assuming (1) the effective field is much smaller than the externally applied field and (2) both the in-plane magnetic and shape anisotropy of the FM layer can be neglected. The former allows one to evaluate the influence of effective field using first order approximation whereas the latter makes it possible to decompose the effective field into different components and calculate their effects on the magnetization separately[18-20]. The first assumption is based on the fact that the effective fields at the current density used in our experiments are at least one order of magnitude smaller than the external field, and the second assumption is due to the actual sample geometry used in the experiment. Based on these assumptions, the Hall voltage can be written as $V_{xy} = V_{xy}(\vec{H}_{ex}) + \Delta V_{xy}(\vec{H}_I)$, where $\vec{H}_I = \vec{H}_{FL} + \vec{H}_{DL} + \vec{H}_{Oe}$ is the sum of SOT effective fields and Oersted field, and $\vec{H}_{ex}$ is the external field applied in the sample plane. Without loss of generality, we assume that the applied field points in the direction defined by polar angle $\theta_H$ and azimuthal angle $\varphi_H$, i.e., $\vec{H}_{ex} = H_{ex}(\sin\theta_H \cos\varphi_H, \sin\theta_H \sin\varphi_H, \cos\theta_H)$, and at the equilibrium state, the magnetization points in the direction of $\hat{m} = $ $(\sin\theta_m \cos\varphi_m, \sin\theta_m \sin\varphi_m, \cos\theta_m)$, where $\theta_m$ and $\varphi_m$ are the polar and azimuthal angle of the magnetization, respectively. Then, the corresponding Hall voltage may be written as

$$V_{xy}(\vec{H}_{ex}) = V_{AHE0}\cos\theta_m + V_{PHE0}\sin^2\theta_m \sin(2\varphi_m), \tag{1}$$

where $V_{AHE0}$ and $V_{PHE0}$ are the half peak-to-peak voltage of AHE and PHE, respectively. Since the current induced field is small, its contribution to Hall voltage can be approximated as

$$\begin{aligned}\Delta V_{xy}(\vec{H}_I) &= \frac{dV_{xy}(\vec{H}_{ex})}{d\theta_m}\Delta\theta_m + \frac{dV_{xy}(\vec{H}_{ex})}{d\varphi_m}\Delta\varphi_m \\ &= [-V_{AHE0}\sin\theta_m + 2V_{PHE0}\sin\theta_m\cos\theta_m \sin(2\varphi_m)]\Delta\theta_m \\ &\quad + 2V_{PHE0}\sin^2\theta_m \cos(2\varphi_m)\Delta\varphi_m,\end{aligned} \tag{2}$$

where $\Delta\theta_m$ and $\Delta\varphi_m$ are the change in polar and azimuthal angle of the magnetization caused by the current induced field.

In the present case, since a symmetrical Hall cross is used, the in-plane shape anisotropy in the central cross region is negligibly small, and therefore, we can assume that $\theta_m \approx \theta_H = \pi/2$ and $\varphi_m \approx \varphi_H$ when the applied field is in the $xy$ plane. Any deviation of $\theta_m$ from $\pi/2$ is caused by field misalignment. As discussed above, when a current flows in the sample, $\vec{H}_I$ will be induced, which will affect the equilibrium magnetization direction. Since $\vec{H}_I$ is much smaller than $\vec{H}_{ex}$, the current induced change in $\theta_m$ and $\varphi_m$ can be estimated using the first order approximation[18-19], which gives $\Delta\theta_m \approx \frac{H_I^\theta}{H_{ex}+H_d}$ and $\Delta\varphi_m \approx \frac{H_I^\varphi}{H_{ex}}$; here, $H_I^\theta$ ($H_I^\varphi$) is the $\theta$ ($\varphi$) component of $\vec{H}_I$, $H_d$ is the demagnetizing field in $z$ direction. When $\theta_m \approx \pi/2$, $H_I^\theta$ is mainly contributed by the DL effective field, and therefore, $H_I^\theta = -H_{DL}[\hat{m} \times (\hat{z} \times \hat{\jmath})]_z = -H_{DL}\cos\varphi_H$; here, the subscript $z$ indicates the $z$ component. On the other hand, $H_I^\varphi$ is given by the projection of $\vec{H}_{FL} + \vec{H}_{Oe}$ in the azimuthal direction, *i.e.*, $H_I^\varphi = (H_{FL} + H_{Oe})\cos\varphi_H$. Substituting all these parameters into Eq. (2), we obtain

$$\Delta V_{xy}(\vec{H}_I) = \left(V_{AHE0}\frac{H_{DL}}{H_{ex}+H_d}\right)\cos\varphi_H + \left(V_{PHE0}\frac{H_{FL}+H_{Oe}}{H_{ex}}\right)(\cos\varphi_H + \cos 3\varphi_H). \tag{3}$$

It is apparent that, the effective field induced Hall signal has two components: one is due to AHE and the other is related to PHE. Figs. 1(b) and 1(c) show the normalized AHE and PHE signal as a function of $\varphi_H$. As we will discuss shortly, the PHE contribution is typically one order of magnitude smaller than that of AHE; therefore, it can be subtracted out from the overall output signal by treating its amplitude as a fitting parameter. Once the PHE component is removed, the output signal will be proportional to $cos\varphi_H$; therefore, the Hall cross functions effectively as an angle sensor. The advantage of using SOT Hall device is that, any offset in the device can be readily compensated by measuring the Hall voltage using two consecutive pulses with opposite polarity and then adding up the two measurement results to generate the output signal. This is possible because $\Delta V_{xy}(\vec{H}_I)$ is an even function, whereas the offset voltage $V_{xy}(\vec{H}_{ex})$ is an odd function of the driving current. The output signal is simply given by $2\Delta V_{xy}(\vec{H}_I)$.

We now turn to the fabrication and evaluation of SOT-based angle sensors. In order to resolve the full range of 360º, we employed two identical Hall crosses with their current axes aligned orthogonal to each other, as shown schematically in Fig. 2. The two crosses, with a dimension of 5 μm (width) × 50 μm (length), are placed as close as possible on a same substrate and fabricated in a same experimental run. To facilitate discussion, we name them as sensor A and B, respectively. The Hall cross stack, comprising of Pt(3)/Co(3), was deposited on $SiO_2$/Si substrate by magnetron sputtering with a base pressure of $2 \times 10^{-8}$ Torr and working pressure of $3 \times 10^{-3}$ Torr. Both the current and voltage electrodes were formed by Pt(10)/Cu(200)/Ta(5). The number inside the parentheses is thickness in *nm*. As indicated in Fig. 2, a rotating in-plane magnetic field of 250 – 2000 Oe was applied to mimic actual angle detection setup. Current pulse ($I_{pulse}$) is injected in *x* direction for sensor A and *y* direction for sensor B. Meanwhile, for each polarity of current pulse, the transverse voltage is recorded as $V_+$ for positive current and $V_-$ for negative current. The pulse width of $I_{pulse}$ is 5 $ms$. The measurements were repeated for ten times at each angle position to obtain an average voltage signal with low noise. The differential voltage signal $\Delta V = |V_+| - |V_-|$ is then extracted as a function of the direction of in-plane magnetic field, ranging from 0° to 360°.

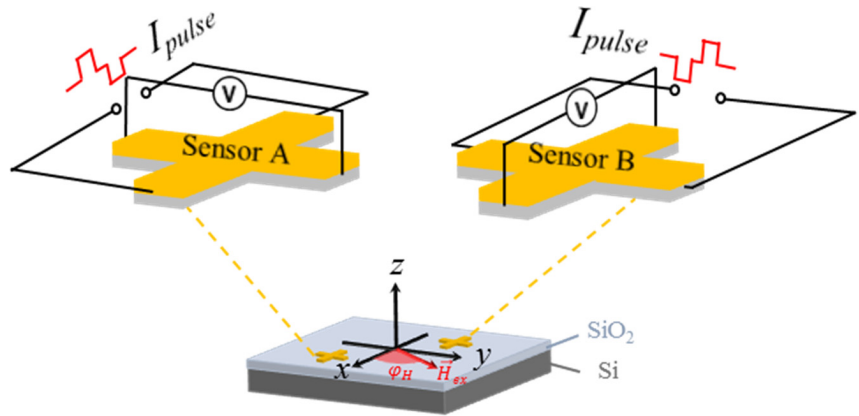

FIG. 2. Measurement geometry for evaluating the performance of angular position sensor comprised of two identical Hall crosses formed on a same substrate. The Hall crosses are oriented such that the current directions are orthogonal to each other.

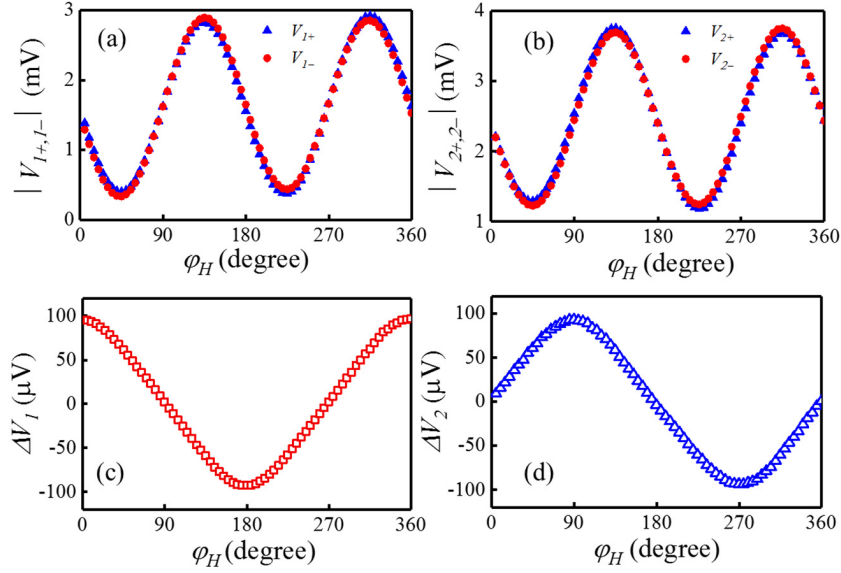

FIG. 3. (a)-(b) Absolute values of Hall voltage signal obtained at ± 5 mA for sensor A and B as a function of in-plane field (500 Oe) angle $\varphi_H$. (c)-(d) Differential Hall voltage for sensor A ($\Delta V_1$) and B ($\Delta V_2$).

Figure 3(a) and 3(b) show the absolute values of Hall voltage obtained at ± 5 mA as a function of rotating field angle for sensor A and B, respectively. The amplitude of the field in this specific measurement is 500 Oe. As expected from Eq. (1), the Hall signal is dominated by PHE and shows an angle dependence close to $\sin 2\varphi_H$. This is understandable because the magnetization direction follows closely the external field direction and rotates in the plane, *i.e.*, $\theta_m \approx \theta_H = \pi/2$ and $\varphi_m \approx \varphi_H$. Despite its large amplitude, the PHE signal cannot be used to calculate the field angle as it is because it does not follow exactly the sinusoidal waveform. In order to extract the useful signal, we focus on the small difference between the output signal obtained by the positive (filled triangle) and negative (filled circle) current pulse shown in Figs. 3(a) and (b). The corresponding differential voltage signals ($\Delta V_1$, $\Delta V_2$) are shown in Fig. 3(c) and Fig. 3(d), respectively. As can be seen from the figures, $\Delta V_1$ shows nearly a cosine whereas $\Delta V_2$ exhibits a sine shape, resulting from the orthogonal alignment of the current axes for the two Hall crosses. The slight deviation from the exact cosine or sine shape is due to the fact that the measured differential Hall voltage contains both AHE and PHE contributions, as described by Eq. (3). In order to obtain pure cosine and sine signals so as to calculate the field angle, we have to subtract out the PHE contribution from the measured signals. As the absolute value of PHE contribution is unknown, we perform the curve fitting on $\Delta V_1$ using the following equation:

$$\Delta V_1 = C_{AHE1} \cos(\varphi_H - \varphi_{01}) + C_{PHE1}[\cos(\varphi_H - \varphi_{01}) + \cos3(\varphi_H - \varphi_{01})] + C_{01}, \quad (4)$$

where $C_{AHE1}$ and $C_{PHE1}$ are the coefficient of AHE and PHE component, $\varphi_{01}$ and $C_{01}$ are the angle and voltage offset, respectively. After the best fitting is obtained, we can subtract out the PHE contribution to obtain $\cos(\varphi_H - \varphi_{01})$, which is given by

$$\cos(\varphi_H - \varphi_{01}) = \{\Delta V_1 - C_{PHE1}[\cos(\varphi_H - \varphi_{01}) + \cos3(\varphi_H - \varphi_{01})] - C_{01}\}/C_{AHE1}. \quad (5)$$

Similarly, from the output of sensor B, we can extract $\sin(\varphi_H - \varphi_{02})$, which reads

$$\sin(\varphi_H - \varphi_{02}) = \{\Delta V_2 - C_{PHE2}[\sin(\varphi_H - \varphi_{02}) - \sin3(\varphi_H - \varphi_{02})] - C_{02}\}/C_{AHE2}. \quad (6)$$

The best fitting of the curves shown in Figs. 3(c) and 3(d) yields $\varphi_{01}(\varphi_{02}) = 0.61°(1.13°)$, $C_{AHE1}(C_{AHE2}) = 8.56(8.49) \times 10^{-5} V$, $C_{PHE1}(C_{PHE2}) = 4.43(4.19) \times 10^{-6} V$, and $C_{01}(C_{02}) = 1.98(0.56) \times 10^{-6} V$. Here, $\varphi_{01}(\varphi_{02})$ is the misalignment of the field direction with respect to the current direction for sensor A and B, respectively. For both sensors, the voltage offset is around two orders of magnitude smaller than the signal, indicating that the pulse measurement effectively suppresses the DC offset. Meanwhile, the coefficient of AHE is one order of magnitude larger than that of PHE, suggesting that AHE is dominant. Figs. 4(a) and (b) show the sum (symbols) of the PHE signal and DC offset, which indeed follows the $\varphi_H$-dependence predicted by Eq. (3) (solid lines). Therefore, in practical applications, one can obtain $C_{PHE}$ and $C_0$ first from pre-calibration measurements and then uses them to subtract out the PHE contribution and DC offset from the measured data. Any deviation in the signal amplitude of the two sensors can also be compensated by using the experimentally extracted $C_{AHE}$ to normalize the sine and cosine output signals.

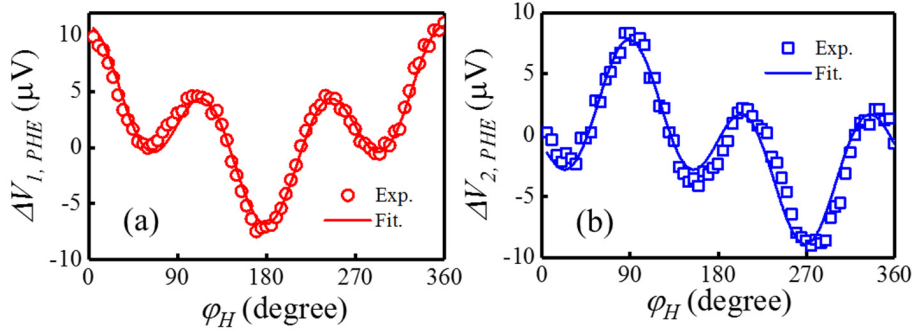

FIG. 4. (a)-(b) Experimental (symbols) and fitting results (solid line) of PHE component for sensor A (a) and B (b), respectively.

The detected field angle can be calculated from the corrected and normalized sine and cosine signal as follows:

$$\varphi = arctan2\,[\sin(\varphi_H - \varphi_{02}), \cos(\varphi_H - \varphi_{01})] + (\varphi_{01} + \varphi_{02})/2. \quad (7)$$

As there is a difference of 0.52º between the offset angle $\varphi_{01}$ and $\varphi_{02}$ due to misalignment of two cross structures during lithography, we add $(\varphi_{01} + \varphi_{02})/2$ as an average offset to obtain the calculated angle $\varphi$. Since arctan2 gives an angle in the range from –180º to + 180º, it is mapped to the range between 0 and 360º, as done in common practice. The calculated angle $\varphi$ as a function of the field position $\varphi_H$ is depicted in Fig. 5(a), from where an almost linear correspondence between the two is observed, suggesting that the calculated value follows closely the actual rotational angle. The angle error, $\varphi - \varphi_H$, at each position is shown in Fig. 5(b). Except for a few positions, the angle error is well below 1º, with an average value of 0.65º.

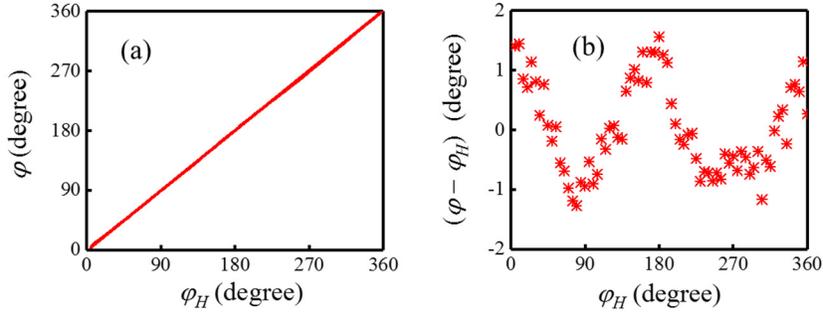

FIG. 5. (a) Measured rotational angle $\varphi$ as a function of field angle $\varphi_H$. (b) Distribution of angle error within the full angle range of 360º.

We have also performed the same measurements and data fitting for $H_{ex}$ = 250, 1000 and 2000 Oe. The maximum, minimum and average angle errors are summarized in Table I. In general, both the maximum and average errors become smaller when the field strength increases. With an applied field above 500 Oe, the angle error is comparable to most commercial GMR and TMR based angle sensors[21-22], though the SOT based device has a much simpler structure. Although the AMR sensor can give a much better accuracy, it requires an additional Hall sensor to differentiate the angle between 0-180º and 180-360º due to its quadratic dependence on the field angle.

TABLE I. Angle errors at different external field.

| Angle error (degree) | External field (Oe) | | | |
|---|---|---|---|---|
| | 250 | 500 | 1000 | 2000 |
| Max. | 2.4188 | 1.5623 | 1.3721 | 1.2168 |
| Min. | 0.0156 | 0.0152 | 0.0113 | 0.0136 |
| Ave. | 1.0598 | 0.6490 | 0.4516 | 0.3795 |

In summary, we have devised an alternative type of angular position sensor based on the SOT effect, which consists of only two Hall crosses. The sinusoidal dependence of the DL effective field on the in-plane field direction generates a sine or cosine waveform from the two orthogonally aligned Hall devices, which are in turn used to calculate the field angle. An average angle error of 0.65º over the 360-degree range is obtained from the prototype device, and it could be further reduced by optimizing the materials and device geometry. Our design can potentially lead to a low cost angular position sensor considering its very simple structure.

This work is supported by the Singapore National Research Foundation, Prime Minister's Office, under its Competitive Research Programme (Grant No. NRF-CRP10-2012-03). Y.H.W. is a member of the Singapore Spintronics Consortium (SG-SPIN).

**FIGURE CAPTIONS:**

FIG. 1. (a) Illustration of the SOT effective fields in a Pt/Co bilayer induced by a current in *x*-direction ($\vec{H}_{ex}$: external field, $\hat{m}$: magnetization direction, $\hat{\sigma}$: spin polarization). (b) and (c): simulated differential AHE (b) and PHE (c) signals as a function of $\varphi_H$ induced by two current pulses with opposite polarity.

FIG. 2. Measurement geometry for evaluating the performance of angular position sensor comprised of two identical Hall crosses formed on a same substrate. The Hall crosses are oriented such that the current directions are orthogonal to each other.

FIG. 3. (a)-(b) Absolute values of Hall voltage signal obtained at ± 5 mA for sensor A and B as a function of in-plane field (500 Oe) angle $\varphi_H$. (c)-(d) Differential Hall voltage for sensor A ($\Delta V_1$) and B ($\Delta V_2$).

FIG. 4. (a)-(b) Experimental (symbols) and fitting results (solid line) of PHE component for sensor A (a) and B (b), respectively.

FIG. 5. (a) Measured rotational angle $\varphi$ as a function of field angle $\varphi_H$. (b) Distribution of angle error within the full angle range of 360°.